\documentclass[twocolumn]{aastex62}
\usepackage{amsmath}
\usepackage{txfonts}
\usepackage{ae,aecompl}
\usepackage{graphicx}
\usepackage{color,soul}

\newcommand{\beq}{\begin{equation}}
\newcommand{\eeq}{\end{equation}}
\newcommand{\bea}{\begin{eqnarray}}
\newcommand{\eea}{\end{eqnarray}}

\newcommand{\pv}{\mathrm v}
\newcommand{\vp}{\varphi}
\newcommand{\rs}{r_{\rm s}}
\newcommand{\rd}{\rm d}

\shorttitle{Planet disruption}
\shortauthors{Jia \& Spruit}

\begin{document}

\title{Disruption of a planet spiralling into its host star}

\correspondingauthor{Shi Jia}
\email{jiashi@ynao.ac.cn}

\author{Shi Jia}
\affiliation{Yunnan Observatories, Chinese Academy of Sciences, Kunming, 650011,China}
\affiliation{Max-Planck-Institut f\"{u}r Astrophysik, Karl-Schwarzschild-Str.\ 1, D-85748 Garching, Germany}
\affiliation{Key Laboratory for the Structure and Evolution of Celestial Objects, Chinese Academy of Sciences, Kunming 650011, China}
\affiliation{University of Chinese Academy of Sciences, Beijing 100049, China}

\author{H.C.\ Spruit}
\affiliation{Max-Planck-Institut f\"{u}r Astrophysik, Karl-Schwarzschild-Str.\ 1, D-85748 Garching, Germany}

\begin{abstract}

The processes leading deformation and destruction of planets spiraling into the convective envelope of their host stars are described. The planet is compressed by the ram pressure, and deformed into a flattened shape for which a quantitative  model is developed. Compression increases the planet's density contrast with the envelope and its gravitational binding energy. This increases the survivability especially of gas planets. An estimate is given for the depth of disruption by ram pressure, and for the subsequent fragmentation of the remnants. We show how the debris of rocky or iron planets, instead of mixing through the convection zone, sinks below the base of the convection zone. The time scale of the entire sequence of events is of the order of a few orbital times of the planet. If spiral-in of (partly) icy, rocky or iron planets has happened to the pre-main sequence Sun, it could account for the higher opacity below the base of the convection zone as inferred from helioseismology.

\end{abstract}


\section{Introduction}

A correlation between the presence of a gas giant planet and its host stellar metallicity has been well established over the last two decades (e.g., Gonzalez 1997; Santos et al. 2004; Fischer \& Valenti 2005; Johnson et al. 2010), although there is still a debate for the planet-metallicity correlation of Neptune size and smaller planets (e.g., Wang \& Fischer 2015; Schuler et al. 2015, and references therein). Two main scenarios have been proposed to explain the planet-metallicity correlation. The primordial hypothesis (Pinsonneault et al. 2001) assumes that stars with planets are formed from metal-rich clouds. In the core accretion model of gas giant planet formation (Pollack et al. 1996), this hypothesis implies that the star forms metal-rich as a whole. The second scenario, the inhomogeneous accretion hypothesis, assumes that the higher metallicity of planet hosting stars derives from the accretion of material of enhanced metallicity during later stages of the star formation process (e.g., Gonzalez 1997; Laughlin \& Adams 1997; Murray et al. 2001). In this scenario, metal enhancement only occurs in the stellar convective envelope. The observed abundances should then decrease with increasing depth of the convective envelope. Such a correlation is not seen in the observations (e.g., Pinsonneault et al. 2001; Santos et al. 2001; Fischer \& Valenti 2005). The scenario also depends on the assumption that the enhancement remains confined to the convective envelope for much of the life of the star. This would not be happen if metal rich matter can settle into the radiative interior by thermohaline mixing (Vauclair 2004; Denissenkov \& Merryfield 2011; Th{\'e}ado \& Vauclair 2012). Depending on the (rather uncertain) efficiency of this process, it would even out metallicity differences between envelope and interior. 

The inhomogenous accretion scenario requires that material accreting later in the star formation process is enhanced in metals. This could be the case if this material derives from planets migrating to their host star by interaction with the accretion disk (e.g., Lin 1997; Laughlin \& Adams 1997; Sandquist et al. 1998, 2002). Whether this leads to enhanced abundances at the surface of the star depends on how mass transfer from the planet to the star takes place at the end of the migration (Sandquist et al. 1998, 2002; Th{\'e}ado \& Vauclair 2012). Depending on the mass and radius of planet and that of host and the equation of state of planet, transfer can be a slow process (in the case of stable Roche lobe overflow), rapid in the case of dynamically unstable Roche lobe overflow, or the planet can enter the star whole before Roche lobe overflow takes place (the `direct merger' case, see Jia \& Spruit 2017 for a recent analysis and references therein).

An important issue is whether (much of) a planet can survive its travel through the convection zone (CZ), and dump its metal load on the radiation interior. The surface metallicity enhancement would then be negligible. Numerical simulations of direct merger by Sandquist et al.\ (1998, 2002) suggested that planets dissolve only gradually while spiraling in through the convective envelope of a sun-like star, but in some cases survive till the base of the CZ.

Survival of planets spiraling in is made possible by their gravitational binding energy, as well as the high density of the planet's envelope relative to the star's CZ. These conditions also present obstacles to realistic numerical simulations, which are affected additionally by numerical diffusion (of momentum and of heat) due to finite spatial resolution. In the following we study the problem with a more analytical approach, addressing separately the processes of ablation of the outer layers of a planet, of the distortion of its shape under the ram pressure of a supersonic flow, and of its final disruption when the increasing ram pressure approaches its gravitational binding energy. The focus is on the depth the planet reaches before it is disrupted and on the fate of its debris. 

For a planet to enter its host star, it has to survive mass loss by Roche lobe overflow before it touches the surface of its host star. For main sequence stars this is the case only for iron-dominated or rather massive rocky or giant planets. Spiral-in is most likely to happen early in the evolution of the star, when its radius is larger and its mean density much lower than on the main sequence (Jia \& Spruit 2017). For the examples given below, we assume a $1 M_\odot$ star at a nominal age of 3.8 Myr, when its radius is $1.5 R_\odot$, its mean density $0.41 \ \mathrm{g \ {cm}^{-3}}$. 

\section{Spiral-in and disruption}
\label{spiraldiss}

The planet is moving with a supersonic velocity relative to the ambient stellar gas forming a shock front. Its velocity is reduced by the associated drag force. As the planet encounters the stellar surface, the drag force is small because of the low density of stellar atmosphere. The orbit is initially still close to Keplerian (for a theoretical analysis see Metzger et al. 2012).  Once the planet is completely engulfed in the stellar envelope, the drag force is  much larger. The spiral-in time scale for a main sequence (MS) host star can be on the order of the orbital period ($\sim 10^4$ s) (Sandquist et al. 1998, 2002). Figure\ \ref{spiral-velocity} shows the path of a 20 $M_\oplus$ iron planet orbiting into the convective envelope of a $1 M_\odot$ star (solar metallicity) with different stellar ages. In evolved stars, however, the process can take many orbits. An example for a moderately evolved $1\,M_\odot$ subgiant is shown in panel C of Figure\ \ref{spiral-velocity}. The spiral-in time scale in a red giant host can be on the order of thousands of years (e.g.,\ Livio \& Soker 1984).

For the calculations of Figure\ \ref{spiral-velocity}, the planets are assumed to remain intact during the spiral-in, at their original mass. In reality, the mass will change in the process. The planet gets compressed by the ram pressure and external gas pressure. Interaction with the flow may cause mass loss by erosion (ablation) of the planet's surface. In Section \ref{ablat} we will argue that ablation may contribute, but is probably secondary to fragmentation by a global process as discussed below and in Section \ref{disruption}.

On the short spiral-in time scale, compared with the thermal time scale of the planet interior, compression by the external pressure takes place adiabatically. It can be computed approximately with polytropic models of fixed entropy, or from more realistic interior models. This is the subject of Section \ref{compress}. 

The orbital velocity of the planet inside the host star is highly (initially) to mildly (later) supersonic.  The difference in pressure between the side facing the incoming flow (with velocity $v$) and the backside is of the order of the ram pressure, $\rho_{\rm ext}{\pv}^2$, where $\rho_\mathrm{ext}$ is the density of stellar material surrounding the planet. This difference acts as a drag force on the planet, but also distorts its shape, flattening it in the flow direction. A model for this distortion of the planet is developed in Section \ref{distortion}.  

\subsection{Compression}
\label{compress}
One might think that a gas planet, with its low mean density, and even lower density of its envelope, would be quickly stripped and dissolved when it enters a stellar envelope. This is probably not the case, as we argue in the following.  

A Jupiter near a main-sequence host star would dissolve by Roche lobe overflow before touching the stellar surface. For such host stars spiral-in is relevant only for planets with a higher mean density. If the host is a pre-main sequence star (PMS) or a moderately evolved star, on the other hand, even a planet with the mean density of a gas giant will touch its surface before Roche lobe overflow. From then on, the planet finds itself in a high pressure environment, compressing its outer layers. 

At the depth in Jupiter where the pressure is $2$ Mbar ($2 \times 10^{12} \ \mathrm{g\ {cm}^{-1} \ s^{-2}}$), for example, the density is $\sim 1 \ \mathrm{g \ {cm}^{-3}}$(cf. the review in Militzer et al. 2016). In  the star's envelope the density is much smaller at the same pressure, especially in an evolved star.  The host star in panel C of Figure\ \ref{spiral-velocity}, for example, the density at 2 Mbar is $\sim 0.001 \ \mathrm{g \ {cm}^{-3}}$. In the envelope of a main sequence star like the Sun it is about $\sim 0.03 \ \mathrm{g \ {cm}^{-3}}$.
Assuming that the compression takes place adiabatically, there is a simple estimate for the density ratio between the planet's surface and its environment, if both the planet interior and the convective envelope around it are stratified nearly adiabatically (isentropically). If $\gamma$ is the ratio of specific heats (assumed constant), the pressure $p$ and density $\rho$ are related by
\beq P=K \rho^\gamma,\eeq
where the constant $K$ is related to the value of the entropy. At a given external pressure $P_{\rm ext}$, pressure equilibrium at the planet's surface, $P_ {\rm ext}=P_{\rm surf}$ yields (assuming $\gamma$ the same in planet and the CZ of the star):
\beq s\equiv {\rho_{\rm ext}\over\rho_{\rm surf}}=\left({K_{\rm surf}\over K_{\rm ext}}\right)^{1 \over \gamma}, \label{sdens}\eeq
where $\rho_{\rm ext}$ is the density of the stellar environment, $\rho_{\rm surf}$ is the surface density of the planet, $K_{\rm surf}$ and $K_{\rm ext}$ are the K for the planet interior and the stellar envelope, respectively. This estimate of the density ratio $s$ between planet's surface and its environment is independent of the depth in the stellar envelope, around a factor 0.001 for the evolved star and cold planet as mentioned above. The estimate has its limitations because the pressure-density relation of matter in a planet is different from that of an ionized ideal gas. The isentropic  pressure-density relation for the degeneracy-supported interiors of gas giant planets corresponds approximately to a $\gamma$ of 2 (e.g., Stevenson 1982; Militzer et al. 2016). This is in fact not too different from the value $\gamma\approx 5/3$ that holds for the equation of state as well as for the stratification of a convective stellar envelope.

For substantially heated planets (`hot Jupiters'), the entropy difference between planet and CZ would be somewhat less, and the surface density somewhat lower than estimated from Equation (\ref{sdens}). More detailed calculations are given in Section \ref{distortion}.
The assumption of adiabatic compression made in the above raises the question how long a planet can survive in the environment of a stellar interior, at $\sim 10^6$ K. Heating of the planet's surface is limited by the rate of thermal diffusion from its surroundings. As discussed in Section \ref{ablat} on ablation, the effect is small for the short duration of the spiral-in process.

\subsection{Accretion}
\label{accret}
It has been argued that the planet may be able to accrete mass from its host during the spiral-in. Bondi's (1952) spherical accretion result has been invoked for such accretion (e.g.,\ Livio \& Soker 1984). This model was developed for accretion onto compact objects; it assumes a radial, steady, adiabatic, non-dissipative flow of gas of constant ratio of specific heats $\gamma$, accreting from infinity. Its main result is that steady accreting flows exist only for $\gamma<5/3$. Approaching the accretor, the temperature $T$ in the flow increases by adiabatic compression. If $\gamma<5/3$, $T$ increases more slowly than the virial temperature $T_{\rm vir}= GM \bar\mu/k_{\rm B} r$, where $r$ is the distance to the accretor, $k_{\rm B} = 1.38 \times 10^{-16} {\rm erg/K}$ is the Boltzmann’s constant, and $\bar\mu$ is the mean molecular weight. Since the pressure scale height $H$ in the atmosphere of the accretor, as a fraction of its radius $R$, is of the order $ T/T_{\rm vir}$, one concludes that an accreting mass can be accommodated in a thin layer on the accretor, provided $\gamma<5/3$. As is the case in the partially ionized gas in a convective envelope. 

This model is not applicable to the case of a planet orbiting in a stellar envelope, however. It assumes accretion from a large distance; the temperature of the flow is then independent of conditions at infinity (within plausible limits). The orbiting planet, however, does not accrete from infinity. It is embedded in a hot environment at the post-shock temperature. At large Mach numbers, the post-shock sound speed is of the order of the incoming flow speed, $c_{\rm s,ext}\sim \pv$. Compare this with the escape speed $\pv_{\rm esc,p}$ from the surface of the planet, 
$\pv_{\rm esc,p}^2=GM_{\rm p}/R_{\rm p}$, where $M_p$ and $R_p$ are the mass and radius of the planet, respectively. With $\pv$ of the order of the orbital speed $\pv^2\sim GM_*/r$:
\beq {T_{\rm ext}\over T_{\rm vir,p}}\sim{c_{\rm s,ext}^2\over \pv_{\rm esc,p}^2}\approx {M_*\over M_{\rm p}}{R_{\rm p} \over r},\label{ttv}\eeq
where $M_*$ is the mass of host star, $r$ is the distance from the planet to the center of the host star. For a Jupiter-like planet orbiting at $1 R_\odot$ in a $1\, M_\odot$ star, this yields $T_{\rm ext}/T_{\rm vir,p}\sim 100$. At this temperature the planet does not accrete. It just sits embedded in a hot environment of which the density is affected only marginally by the planet's gravity.

\subsection{Disruption}
\label{disruption}

Distortion of the planet increases as the planet enters denser regions, eventually leading to breakup. In the following, we define \emph {disruption} as the first stage of breakup, resulting from the development of a global oscillation mode. This is distinguished from the subsequent breakup into many smaller bits which we refer to as  \emph{fragmentation}. This process is discussed in Section \ref{fragm}.

Disruption is expected to happen when the energy involved in distorting the planet, the ram pressure integrated over the cross-section of the planet, approaches its binding energy $E_{\rm b}$ (equal to minus one-half of the gravitational energy). For a polytrope of index 1.5, $E_{\rm b}\approx GM^2/R$ (cf. the lecture notes in Glatzmaier 2013). An estimate of the onset of disruption is thus 
\beq f_{\rm d}\equiv {\rho_{\rm ext}{\pv}^2\over e_{\rm b}}\approx 1,\label{fd}\eeq
where $e_{\rm b} \approx \bar\rho_{\rm p}{\pv}_{\rm esc,p}^2$ is the volume average of $E_{\rm b}$, ${\bar\rho}_p$ is the mean density of the planet. This estimate is analogous to the condition for disruption of liquid droplets moving at high speed in air (see the reviews in Lin \& Reitz 1998, Kim \& Hermanson 2012). In that case the equivalent of the gravitational binding energy in Equation (\ref{fd}) is the energy of surface tension. Equation (\ref{fd}) yields as approximate condition for disruption:
\beq \rho_{\rm ext} \approx {\bar\rho}_p { \pv_{\rm esc,p}^2\over \pv^2} \ \mathrm{or} \ f \equiv \mathrm{\frac{\rho_{ext} v^2}{\bar{\rho}_p v^2_{esc}}} \approx 1.\label{fd2}\eeq

The velocity $\pv$ at the time of disruption is of the order of the velocity $(GM_*/r)^{1/2}$ of a freely orbiting planet, even if it becomes increasingly radial, because the free fall speed at distance $r$ is similar to the orbital velocity (cf. Figure\ \ref{spiral-velocity}, red lines). Then Equation (\ref{fd2}) can also be written as 
\beq 
{\rho_{\rm ext} \over {\bar\rho}_p} \approx {M_{\rm p}\over M_{\rm d}}{r_{\rm d}\over R_{\rm p}}={{\bar\rho}_p \over {\bar\rho}_d}{R_{\rm p}^2\over r_{\rm d}^2},\label{rerb}
\eeq 
where $r_{\rm d}$ is the distance from the center of the star where the disruption takes place, and $M_{\rm d}$, $\bar\rho_{\rm d}$ are the mass and mean density of the star inside the radius $r_{\rm d}$.

The time scale $t_{\rm d}$ for the initial disruption (splitting in two) scales with the dynamical time scale of the planet: 
\beq t_{\rm d} \approx \left(G M_{\rm p} \over R_{\rm p}^3 \right)^{-{1 \over 2}}\approx (G {\bar\rho}_p)^{-{1 \over 2}}.\label{tdis1}\eeq 
Comparing this with the orbital time scale at distance $r_{\rm d}$:
\beq t_{\rm orb} \approx {r_{\rm d} \over \pv} \approx \left({{G M_{\rm d}} \over r_{\rm d}^3}\right)^{-{1 \over 2}} \approx (G {\bar\rho}_d)^{-{1 \over 2}},\label{torb} \eeq
shows that
\beq t_{\rm d}= t_{\rm orb} \left({\bar\rho_{\rm d} \over {\bar\rho}_p} \right)^{1\over2}.\label{tdis}\eeq
Since the planet must have avoided Roche lobe overflow in order to spiral into the star, $\bar\rho_*/ {\bar\rho_p}<1$, where ${\bar\rho}_*$ is the mean density of the host star. If disruption takes place in the outer parts of the star, this shows that the disruption time is less than the orbital time scale, though not by a large factor.

From the deceleration $\dot\pv=F_{\rm drag}/M_{\rm p}$, the time scale for the planet to lose its momentum, the drag time is
\beq t_{\rm drag} = {\pv \over {\dot{\pv}}} \approx {M_{\rm p} \over {\pi R_{\rm p}^2\rho_{\rm ext}\pv}} \approx t_{\rm flo}{{\bar\rho}_p \over \rho_{\rm ext}}=t_{\rm orb}{R_{\rm p}\over r_{\rm d}}{{\bar\rho}_p \over \rho_{\rm ext}},\eeq
where $ t_{\rm flo}=R_{\rm p}/\pv$ is the flow time across the planet, $F_{\rm drag} \approx \pi R_{\rm p}^2\rho_{\rm ext} {\pv}^2$ is the drag force.
With $\rho_{\rm ext}$ from Equation\ (\ref{rerb}) the drag time evaluated at the point of disruption  is
\beq 
t_{\rm drag} \approx t_{\rm orb}{M_{\rm d}\over M_{\rm p}} {R_{\rm p}^2\over r_{\rm d}^2}\approx t_{\rm orb} {g_{\rm d}\over g_{\rm p}},\label{dragd}
\eeq
where $g_{\rm d}$ is the star's acceleration of gravity at $r=r_{\rm d} $, and $g_{\rm p}$ is the planet's surface gravity. The drag time is longer than the orbital time for planets with main sequence hosts, but shorter than the orbit for evolved host stars. 
Comparing the drag time to the disruption time:
\beq  t_{\rm drag} \approx t_{\rm d} {r_{\rm d}\over R_{\rm p}}\left({\bar\rho_{\rm d} \over {\bar\rho}_{\rm p}}\right)^{1\over2}\label{dragdis}.\eeq 
This shows that drag does not affect the planet's velocity much during the disruption process as long as the star is sufficiently close to the main sequence that ${\bar\rho_*/\bar\rho_{\rm p}} > R_{\rm p}^2/R_*^2$, where $R_*$ is the radius of the host star.

\subsection{Ablation}
\label{ablat}
Mass loss by ablation or erosion by the surrounding flow is harder to estimate but potentially relevant. In \emph{thermal ablation}, a layer of some depth at the planet's surface is heated by radiation from the environment, followed by its removal by the surrounding flow. All depends on the hydrodynamics of this removal process. If its time scale $t_{\rm r}$ were known, the depth of the layer that is removed is given by the penetration depth $d_{\rm T}$ of the radiation, $d_{\rm T}\sim (t_{\rm r}\kappa_{\rm T})^{1/2}$, where $\kappa_{\rm T}$ is the thermal diffusivity. The mass loss rate would scale as $\rho_{\rm surf} R_p d_{\rm T} / t_{\rm r}$. Thermal ablation is thus intimately related to the hydrodynamic details of the ablation process. 

The flow around the planet can in principle be stress-free, since viscous interaction is negligible. As long as the surface is smooth, the mechanical effect of the flow is just the ram pressure that decelerates the planet as a whole. The flow is susceptible to Kelvin-Helmholtz instability, however, which produces bumps on the surface with which the flow can interact. The growth rate $\eta$ of this instability depends on the flow velocity $\pv$, the wavenumber $k$, and the density ratio between the environment and the surface of the planet, $s=\rho_{\rm ext}/\rho_{\rm surf}$, for which we have derived an estimate in Section \ \ref{compress}. For $s\ll 1$, the growth rate is\footnote{This result assumes incompressible flow. The post-shock flow surrounding the planet is supersonic with respect to the sound speed of the planet's surface, but subsonic with respect to the temperature in the flow. Corrections for finite Mach number are small in this case (e.g., Fitzpatrick 2017).} (e.g., Fitzpatrick 2017):
\beq \eta=k \pv s^{1 \over 2}, \quad (s\ll 1), \label{gro}\eeq
where we have ignored the planet's acceleration of gravity. This is appropriate for the present case, where the escape speed from the planet is less than the flow speed $\pv$. 

The nonlinear amplitude of the instability determines the thickness of the layer that interacts with the flow. Experiments and observations (e.g., Hwang et al. 2012; Wan et al. 2015) show how the nonlinear development takes the form of `vortex sheet wrapping', producing structures with a thickness $d$ of about half the wavelength  $\lambda$ of the initial disturbance, or $d=\epsilon\, 2\pi/k$, with $\epsilon\approx 0.5$. Since the linear growth rate decreases with wavelength as $1/\lambda$, the thickness of the interacting layer growing from a mixture of modes is dominated by the longest waves that can grow in the available time.  This is the flow time across the planet, of the order $t_{\rm flo} \approx R_p/\pv$. The time  for the mode to grow to its nonlinear state depends on the amplitude of the initial disturbance.  Convective flows in the stellar envelope upstream of the planet are a plausible source of such disturbances, but their amplitude on the length scale of a planet is hard to estimate. Since the initial growth is exponential, however, the dependence on the initial amplitude is modest. We parametrize this by setting
\beq t_{\rm flo}\eta=n,\eeq
where $n$, the number of e-foldings to reach a saturated nonlinear state, will be a modest number, of order $10$, say. Equation (\ref{gro}) then yields 
\beq k R_p s^{1 \over 2}=n,\eeq
and our estimate for the thickness of the layer $d$ becomes, with $\epsilon=0.5$:
\beq {d \over R_p}={\pi\over n}s^{1 \over 2}.\label{dhyd} \eeq
With a thermal diffusivity $\kappa_{\rm T}\sim 10^7$ cm$^2$/s , typical for the deeper layers of a convective envelope, one verifies that the thermal penetration depth $d_T$ over the flow time $\mathrm{t_{flo}}$ is much smaller than $d$, so thermal diffusion does not play a significant role in the ablation process.

With estimate Equation (\ref{dhyd}) for the layer thickness, the mass loss rate by ablation is
\beq \dot m=2\pi R_p\, d\,\, \rho_{\rm surf} \pv_{\rm a},\eeq
where $\rho_{\rm surf}$ is the surface density of the planet (see Section \ref{compress}) and $\pv_{\rm a}$ the velocity with which the ablating layer leaves the planet. The factor $\rho_{\rm surf}\pv_{\rm a}$ is the momentum density of the layer. It acquires this momentum from the external flow through mixing with it. It therefore does not exceed the external momentum density $\rho_{\rm ext}\pv$, but can be of the same order. We set  $\rho_{\rm surf} \pv_{\rm a}\approx \rho_{\rm ext}\pv$, noting that this is probably an overestimate.
The rate of ablation (s$^{-1}$) in units of the planet's mass is then
\beq {\dot m\over M_{\rm p}}= q {\pv\over R_{\rm p}}{\rho_{\rm ext}\over \bar\rho_{\rm p}}s^{1\over2}, \label{dotmM}\eeq
where $q=3\pi\epsilon/n$, an uncertain number of order unity (about 0.5 for the assumptions made in the above).
 
The total mass lost by ablation depends on the duration of the process. The braking force by aerodynamic drag experienced by  the planet  is of the order $F \approx  \pi R_{\rm p}^2 \rho_{\rm ext} \pv^2$, the corresponding deceleration $a=F/M_{\rm p}$ is 
\beq a\approx{\rho_{\rm ext}\over\bar\rho_{\rm p}}{\pv^2\over R_{\rm p}}.\eeq
Under this deceleration, the orbit changes significantly after a time $t_{\rm a}=v/a$:
\beq t_{\rm a}\approx{\bar\rho_{\rm p}\over\rho_{\rm ext}}{R_{\rm p}\over \pv}.\eeq
Having lost much of its angular momentum, the planet then plunges into the interior in a short time, on  a more or less radial orbit, see the example in Figure\ \ref{spiral-velocity}. Comparing with  Equation (\ref{dotmM}) yields a simple result for the mass lost by ablation during the main deceleration phase:
\beq {\Delta m\over M_p}\equiv t_{\rm a}{\dot m\over M_{\rm p}}\approx q\, s^{1/2}.\label{deltam}\eeq
Since the density ratio $s$ is a small number, both for gas giants and rocky planets, we conclude that ablation, though it contributes some, is not the main process destroying the planet. Fragmentation by a global  instability, as discussed in Section \ref{disruption}, is likely to be the main event disrupting the planet. With Equation (\ref{dhyd}), Equation \ (\ref{deltam}) can also be written as 
\beq {\Delta m\over M_p}\approx {d\over R_{\rm p}}.\label{ddm}\eeq
In this form it can be understood by noting that transfer of momentum from the flow to the planet determines both deceleration and ablation. But the cross section for ablation, $2\pi R_{\rm p} d$, is smaller than the cross section of the planet as a whole by a factor of order $d/R_{\rm p}$.

The uncertain coefficient $q$ in Equation (\ref{deltam}) can possibly be calibrated with a suitable numerical simulation. The spatial resolution would have to be sufficient to resolve the thin ablating layer, and able to handle a large density contrast between flow and planet interior. The simulations by Sandquist et al.(2002) were designed for this problem, but it is not clear if numerics can at present be made realistic enough for quantitative measurement of $q$. 

\subsection{First contact}
\label{first}
In the above, we have tacitly assumed that the planet orbits in an environment of homogeneous density. This is not the case if the planet is bigger than the density scale height of the stratification. Near the surface of the star, this is always the case (except if the host is a super giant). The importance of ablation, relative to drag, depends on the relative surface areas facing the incoming flow. In the homogeneous case, we have assumed that the area seen by the drag force is of the order $\pi R_{\rm p}^2$, the area determining ablation is $2\pi R_{\rm p}d$. These areas are different when the planet is not fully engulfed in the star. Interaction then takes place in a reduced area at the interface with the star, the contact surface. For a gas giant (assuming it has not been dissolved by Roche lobe overflow before reaching the surface of its host star) the density of the planet's surface at this contact area is determined by pressure equilibrium, as in the homogeneous case. If the planet has sunk to a depth where the density scale height is $H$, and the area of the contact surface has a width $w$, the areas facing the flow are of the order $w H$ for the drag force and $w d$ for ablation. The effect is that in Equation (\ref{dotmM}) $R_{\rm p}$ is replaced by $H$ if $H<R_p$, so that the ratio of mass loss rate to braking rate is larger than in the homogeneous case:
\beq t_{\rm a}{\dot m\over M}\approx {d\over R_{\rm p} } \, {\rm max}\, (1,{R_{\rm p}\over  H}),\label{ddmH}\eeq
or
\beq t_{\rm a}{\dot m\over M}\approx q\,s^{1/2} \, {\rm max}\, (1,{R_{\rm p}\over  H})\label{ddms}.\eeq

On account of the small density ratio $s$ this still a rather small number, but larger than for a fully immersed planet. This can be understood as a result of the longer braking time, which increases the cumulative effect of ablation.

\section{Calculations}
\label{models}

For a few combinations of host star and planet we calculate how deep into the host star a planet can survive inside stellar the envelope. The stripping of mass (`ablation') has been discussed above (Section \ref{ablat}), where we have found that once a planet has fully entered the host star, ablation is a minor effect during the spiral-in. Ablation may be more important during the first contact with the star (see  Section \ref{first}), when the density scale height of the envelope is still smaller than the size of the planet, as suggested also by Sandquist et al.(2002). 

\subsection{Distortion model}
\label{distortion}

Before it is disrupted by ram pressure, the planet experiences significant deformation, especially on the front side facing the incoming flow. A non-symmetric structure of planet is built up, with the front face compressed and the back side relaxed, flattening the planet in the flow direction\footnote{ Tektites of the australite variety often show such shapes. They are interpreted a having solidified at just the right time to show this deformation (e.g., O'Keefe 1966).}. An accurate calculation of the distortion of the planet is beyond the present scope, but good approximations are possible. 

\begin{figure}
\includegraphics[width=0.45\textwidth]{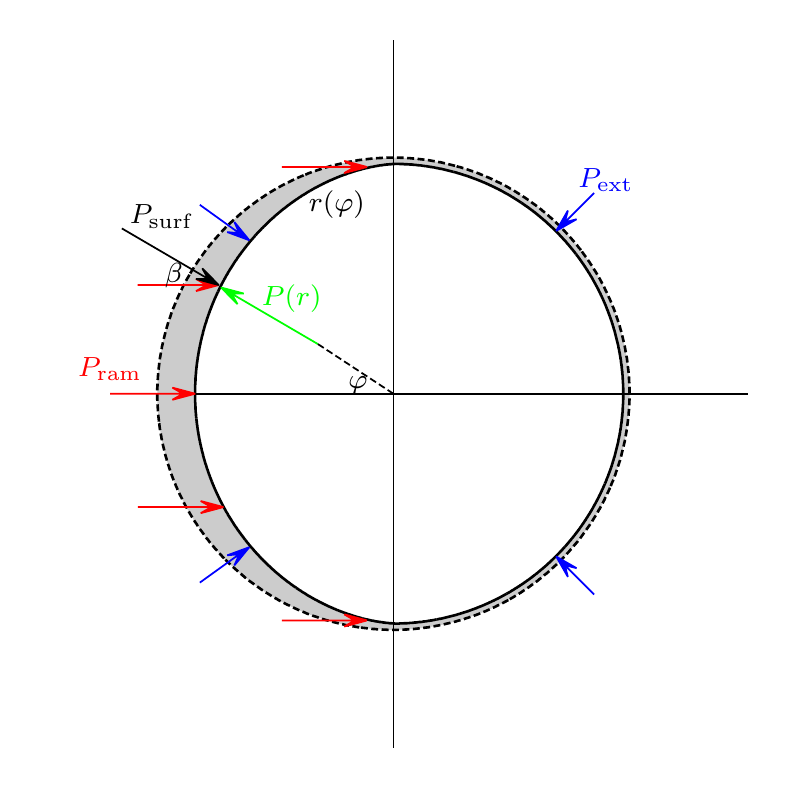}
\caption{Distortion (schematic) of a planet under ram pressure $P_{\mathrm{ram}}$ and external gas pressure $P_{\mathrm{ext}}$ of the external medium. $P_{\mathrm{surf}}$ is the post-shock pressure at latitude $\vp$ [Equation (\ref{ps})]. Solid: the equilibrium surface $r(\vp)$ where $P_{\rm surf}$ matches the internal pressure $P(r)$. Dashed circle is the undistorted planet model.}  
\label{scheme}
\end{figure}

First we need an approximation for the distribution $P_{\rm surf}$ of pressure over the planet's surface. This problem is similar to that of a raindrop moving in air (see, e.g., the treatment by McDonald 1954). The momentum flux tensor of the incoming flow $\bf v$ is  $\rho_{\rm ext}\bf vv$. At large Mach number, the shock surface is close to the planet and the pressure it exerts is the component $\rho_{\rm ext}{\mathrm v}\,{\bf v\cdot n}$, of this tensor, where $\bf n$ is the normal to the (distorted) surface of the planet (cf.\ Figure\ \ref{scheme}). The  pressure which the flow exerts on the planet's surface is then
\beq
P_{\rm surf}(\vp) = \left\{
\begin{array}{ll}
 \rho_{\rm ext}{\mathrm v}^2\,[{\bf \hat v}\cdot {\bf n}(\vp)] + P_{\rm ext}\quad (0<\beta<\pi/2) \\
  P_{\rm ext}\quad (\pi/2<\beta<\pi)
\end{array}
,\right.
\label{ps}
\eeq 
where $\vp$ is the polar angle in spherical coordinates ($r,\vp,\phi$) centered on the planet and with axis along the incoming flow ${\bf v}={\mathrm v}{\bf \hat v}$,  $P_{\rm ext}$ is the pressure of the stellar environment, and $\beta$ is the angle between $\bf {\hat{v}}$ and ${\bf n}(\vp)$. On the front side ($\vp<\pi/2$) this takes into account the angle between the incoming flow and the surface, on the back side it is assumed that only the external pressure remains when the flow has passed over the planet. These assumptions can be relaxed with more realistic models for the post-shock flow, but such elaboration is not warranted in view of a further approximation to be made.

Let $P_{\rm i}(r)$ be the pressure in the undistorted planet as a function of the distance $r$ from its center. As long as $P_{\rm ext}$ is not too large compared with the central pressure of the planet, there exists a surface $\rs(\vp)$ such that [with $P_{\rm surf}$ from Equation (\ref{ps})]:
\beq P_{\rm i}(\rs(\vp))=P_{\rm surf}(\vp).\eeq
Computation of this surface is discussed further below. The mass in the volume extending outside the surface $\rs$ has two effects: its weight determines the pressure it exerts, and the gravitational potential of its mass exerts forces elsewhere in the planet. The latter effect is limited since the mass outside $\rs$  is only a  fraction of the planet  mass. Neglecting these forces, the surface $\rs$ as determined by Equation (\ref{ps}) is the shape of the planet as distorted by the ram pressure, and compressed by  external pressure $P_{\rm ext}$ as well as $P_{\rm ram}$. This is conventionally called a {\em Cowling approximation}: the effect of the gravitational field associated with density perturbations is ignored compared with the other variables (Cowling 1941). It is a good approximation for the oscillation modes of stars.

By omitting the mass $\Delta M$ outside $\rs$, the result  calculated is actually for a planet of somewhat  smaller  mass, $M_{\rm n}=M_0-\Delta M$, where $M_0$ was the mass assumed in the calculation of $\rs$. When a grid of models has been made  as a function of the parameters $M_0$, $\rho_{\rm ext}\pv^2$ and $P_{\rm ext}$, the shape $\rs(\vp)$ for  a fixed mass $M$ is obtained by interpolation in the dependence on $M_{\rm n}$. In the same way, for example, the central pressure $P_{\rm cen}$ can be found for a given value of the planet's mass $M_p$. 

The quantity ${\bf \hat v}\cdot {\bf n}(\vp)$ in Equation (\ref{ps}) can be written in terms of the shape $\rs (\vp)$. With some trigonometry we find
\beq  {\bf \hat v}\cdot {\bf n}= (\sin\vp {1\over \rs}{{\rm d} \rs\over {\rm d}\vp}+\cos\vp)/N,\eeq
where
\beq  N=[1+({1\over \rs} {{\rm d} \rs\over {\rm d}\vp})^2]^{1/2}.\eeq
If $P(r)$ is the internal pressure of the planet as a function of radius $r$, pressure balance at he surface $\rs$ is
\beq   P_{\rm surf}=  P(\rs).\label{pbal}\eeq
To turn Equation (\ref{ps}) into an equation for $\rs$, let $r_{\rm i}(P)$ be the inverse of the function $P(r)$. With Equation (\ref{pbal}), Equation (\ref{ps}) can then be written as
\beq \rs(\vp)=r_{\rm i}(P_{\rm surf}), \eeq
where
\beq
 P_{\rm surf} = \left\{
\begin{array}{ll}
{\rho_{\rm ext}\pv^2}\, (\sin\vp {1\over \rs}{{\rm d} \rs\over {\rm d}\vp}+\cos\vp)/N+ P_{\rm ext} \quad (0<\beta<\pi/2)\\
 P_{\rm ext} \quad (\pi/2<\beta<\pi)
 \end{array}
.\right. 
\eeq 
This is a first order nonlinear ODE for $\rs(\vp)$, with parameters $\rho_{\rm ext}\pv^2$,  $P_{\rm ext}$ and the function $P(r)$. We integrate it with a fourth-order Runge-Kutta method. An example of the results discussed further below is shown in Figure\ \ref{deformation-iron}. The side of planet facing the incoming flow is compressed by the high ram pressure. Without the ram pressure, the back side of the planet is compressed less.

\subsection{Models of planet and star}
\label{mps}
We first consider planet models of uniform composition (iron or rock). A spherically symmetric model of the undistorted planet is obtained by integrating the equations of hydrostatic equilibrium with a fourth-order Runge-Kutta method with adaptive step size (accuracy $ < 10^{-6}$):
\begin{equation}
 \frac{dm}{dr} = 4 \pi r^2 \rho
\end{equation}
\begin{equation}
 \frac{dP}{dr} = \frac{G m \rho}{r^2}
\end{equation}
where r is radius, m is the mass contained in r, $\rho$ is local density of mass shell, P is the pressure, G is the gravitational constant. The equations of state (iron or rock) employed in the integration are the same as used in Seager et al.(2007). Gas giant planets are modeled as polytropes of index $n=1.0$ (Hubbard 1984).

Comparing with the MS stars, close-in planets are easily swallowed by their host stars in the pre-main-sequence period because of the larger stellar radius (Jia \& Spruit 2017). Planet spiral-in process can be effected by the different internal structures between the PMS and MS, therefore, both of them are taken into account during our calculations. The PMS and MS models used were provided by A.\ Weiss (cf. Weiss \& Schlattl 2008).

\subsection{Spiral-in of the planet}
\label{dyp}
We calculate spiral-in trajectories of planets that did not experience Roche lobe overflow before touching the surface of their host stars. In view of the minor importance of ablation as discussed in Sections 2.4 and 2.5, the planet's mass is treated as constant until its dissolution (Section 2.3).
The calculations start at the point when the planet is just immersed as a whole in the host star, so the initial distance from the planet to the center of the host star is $\mathrm{a_{ini} = R_* - R_p}$. The initial orbital period of the planet is defined as $t_{\rm iorb} = \sqrt{GM_*/a_{\rm ini}}$.
We use a polar coordinate system in the plane of the orbit, centered on the center of mass of host star. The drag force is given by:
\begin{equation}
\label{eq:drag}
F_{\mathrm{drag}} = -\frac{1}{2} \rho_{\rm ext} \pv^2 C_{\mathrm{d}} A_{\mathrm{cs}},
\end{equation}
where $\rho_{\rm ext}$ is the density in the host star at the planet's position, 
$\pv$ the velocity of the planet relative to its environment,
$C_{\mathrm{d}}$ is the dimensionless drag coefficient, and
$A_{\mathrm{cs}}=\pi R_{\rm p}^2$ is the planet's cross section. 
Let ${\bf a}(t)=(a,\theta)$ be the position of the planet, ${\bf v}=\rd {\bf a}/\rd t$ its velocity vector, ${\bf \hat v}$ a unit vector in the direction of ${\bf v}$, and $\bf {\hat a}$ a unit vector in the radial direction.
The equation of motion of the planet is then
\beq 
\label{eq:dv}
M_{\rm p}{\rd {\bf v}\over\rd t}=M_{\rm p}{\bf g}+ {\bf F}_{\rm b}\hat{\bf r}-F_{\rm drag}{\bf \hat v},
\eeq
where $\bf g$ is the acceleration of gravity and $F_{\rm b}$ the buoyancy force acting on the planet:
\beq {\bf F}_{\rm b}=- \rho_{\rm ext} V_{\rm p} {\bf g}\eeq
with $V_{\rm p}$ the volume of the planet.
The equation is integrated in time by a straightforward fourth-order Runge-Kutta scheme to yield the planet's distance $a(t)$ from the center of the star. The changes of the planet's shape and volume during the spiral-in process are taken account into the calculations.

\begin{figure*}
\center
\includegraphics[width=0.341\textwidth]{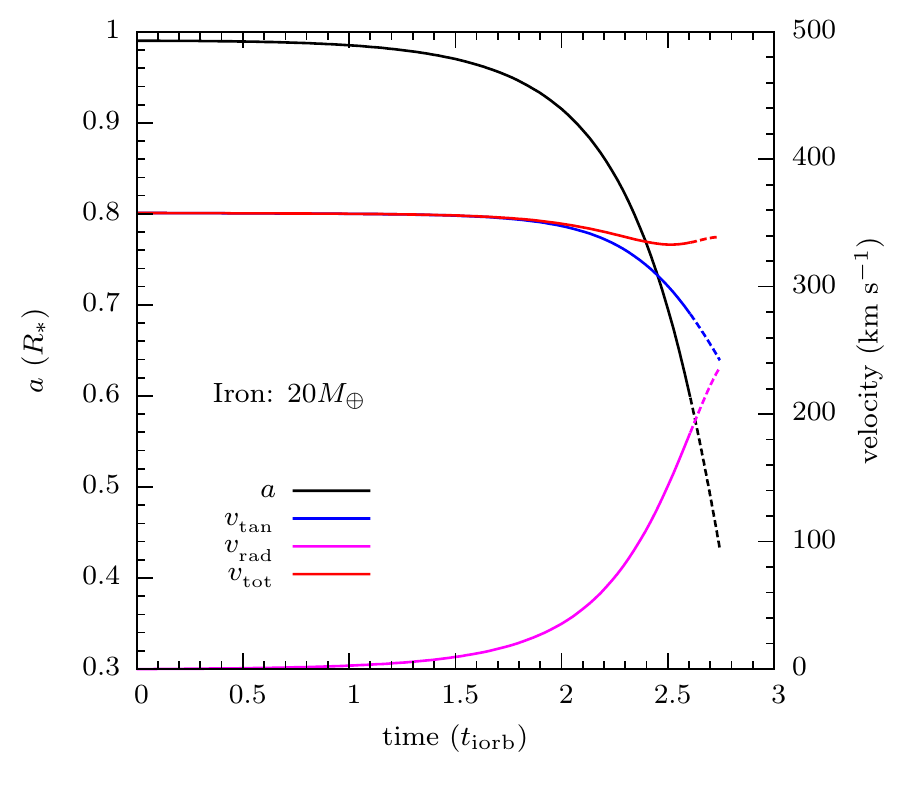}
\includegraphics[width=0.341\textwidth]{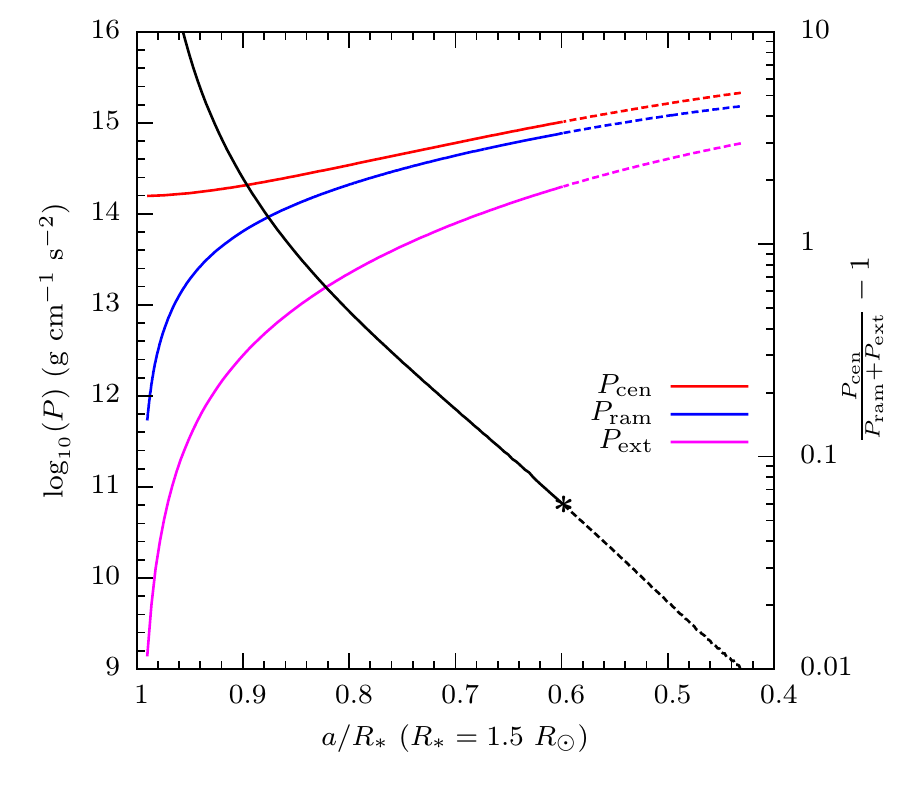}
\includegraphics[width=0.296\textwidth]{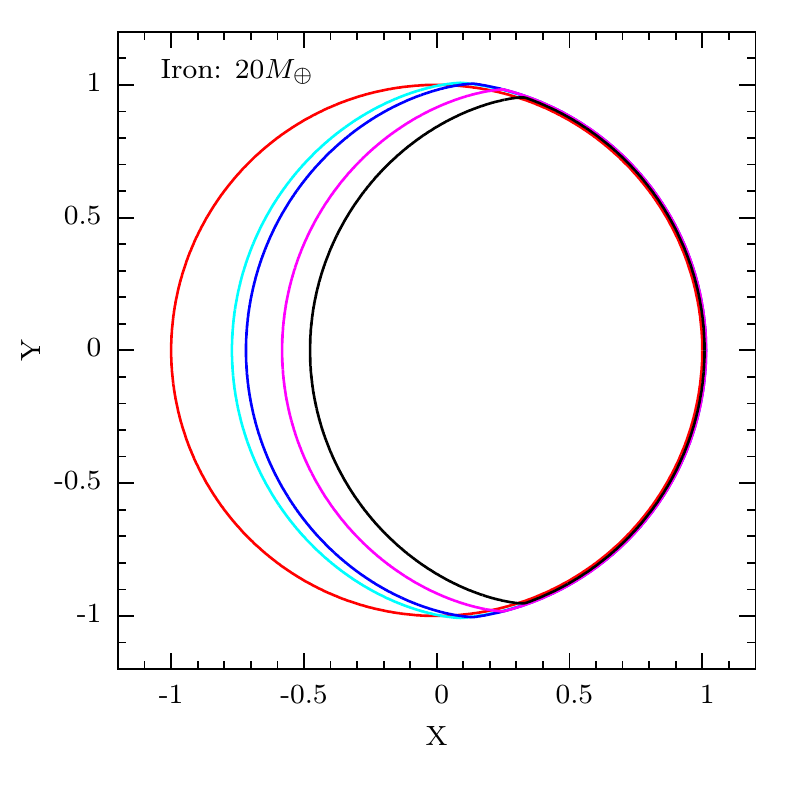}
\caption{A 20 $M_\oplus$ iron planet orbiting in a 1 $M_\odot$ pre-main sequence star with solar matellicity, age 3.8 Myr, and radius 1.5 $R_\odot$. Left: variations of the distance $a$ from planet to the center of the star and the components of the velocity with time (in units of initial orbital period, $t_{\rm iorb} \approx 5 \times 10^3$ s). Middle: variations of the central pressure of the planet, ram pressure and external gas pressure with the distance $a$. The black line shows the relative difference between the central pressure of the planet and the sum of ram pressure and external gas pressure, [$P_{\rm cen}/(P_{\rm ram}+P_{\rm ext})$-1] (right y-axis). The condition for disruption [Equation (\ref{fd2})] is met at $a \approx 0.6$ ($*$). The value assumed for the `disruption factor' $f$, the dimensionless coefficient in the condition for disruption [Equation (\ref{fd2})], is $f=1$. Right: shapes of the distorted planet at increasing depth in the stellar envelope, cyan: $a=0.92 \ R_*$. purple: $a=0.90 \ R_*$. violet: $a=0.85 \ R_*$. black: $a=0.80 \ R_*$ (coordinates in units of the planet's initial radius). Dashed lines (and shapes) are extrapolations beyond the actual disruption depth.}
\label{deformation-iron}
\center
\end{figure*}

\begin{figure*}
\center
\includegraphics[width=0.341\textwidth]{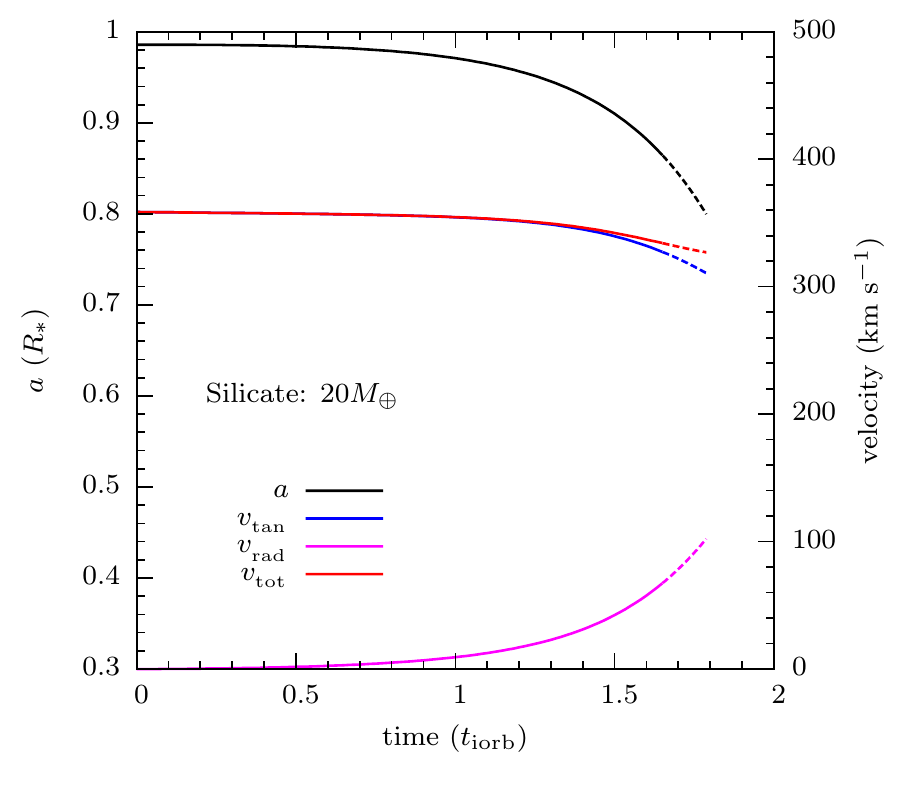}
\includegraphics[width=0.341\textwidth]{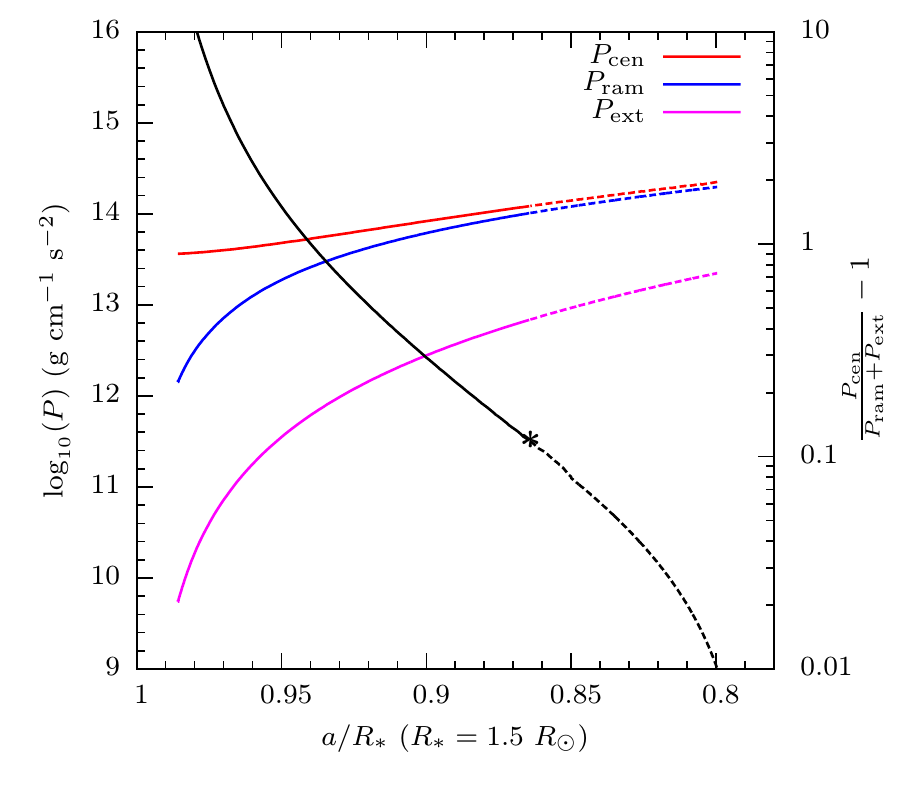}
\includegraphics[width=0.296\textwidth]{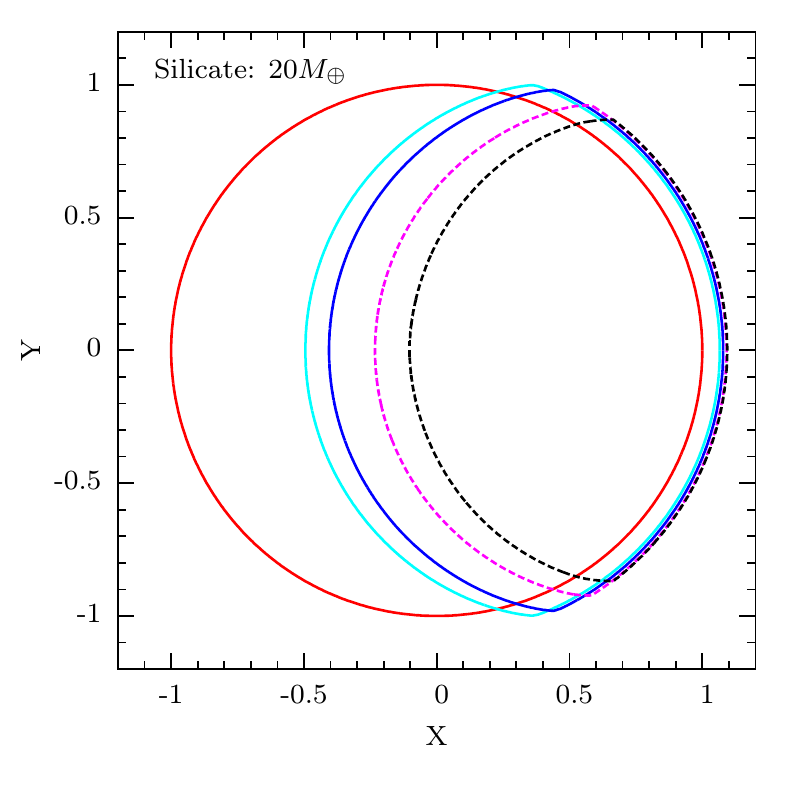}
\caption{Same as Figure \ref{deformation-iron}, but for a rocky planet with 20 $M_\oplus$. Disruption (for assumed disruption factor $f = 1$) occurs at $a \approx 0.86 \ R_*$.}
\label{deformation-rock}
\center
\end{figure*}

\begin{figure*}
\center
\includegraphics[width=0.341\textwidth]{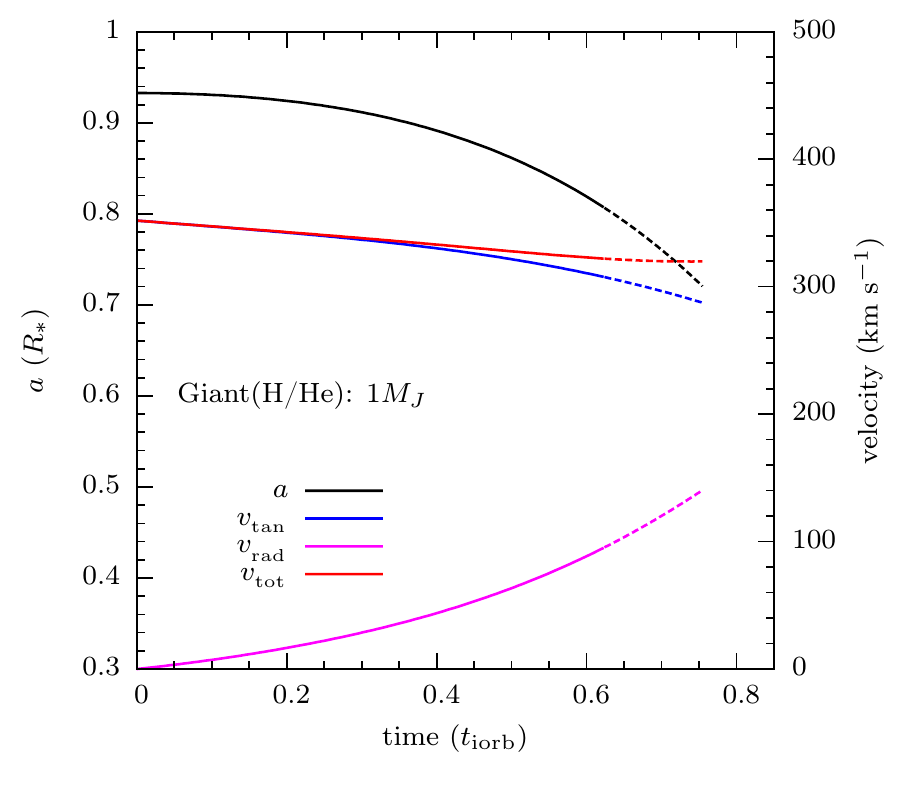}
\includegraphics[width=0.341\textwidth]{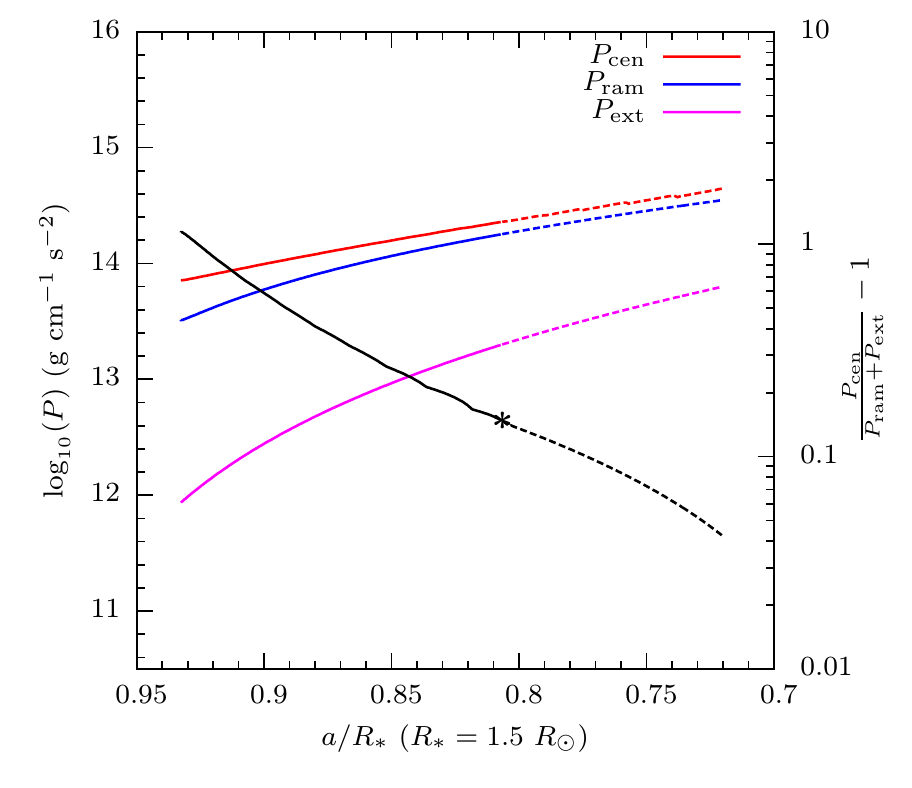}
\includegraphics[width=0.296\textwidth]{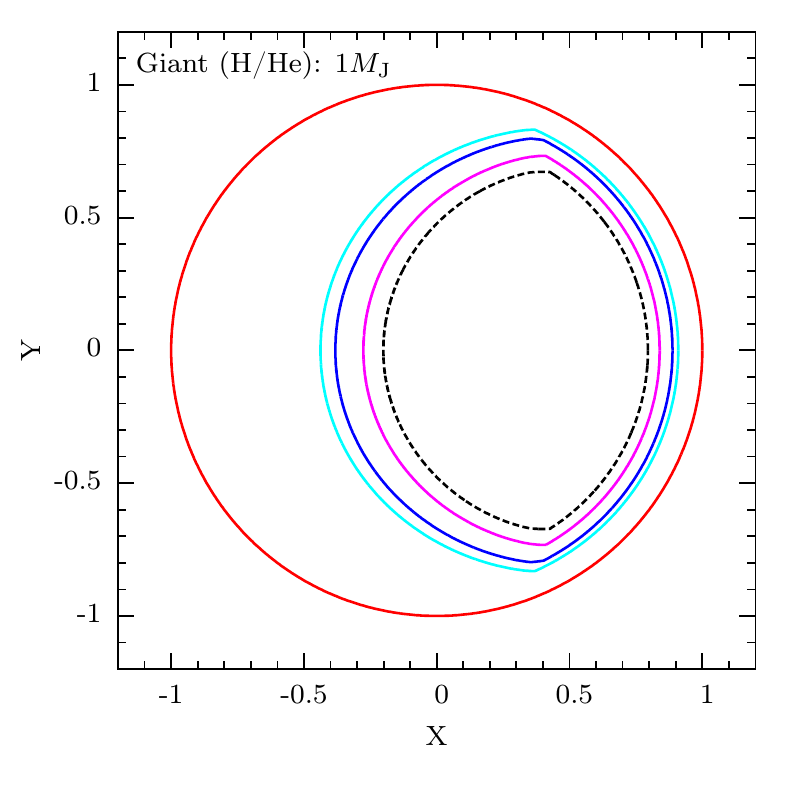}
\caption{Same as Figure \ref{deformation-iron}, but for a giant planet with 1 $M_J$. Disruption (for assumed disruption factor $f = 1$) occurs at $a \approx 0.8 \ R_*$.}
\label{deformation-giant}
\center
\end{figure*}

\begin{figure}
\begin{center}
\includegraphics[width=0.55\textwidth]{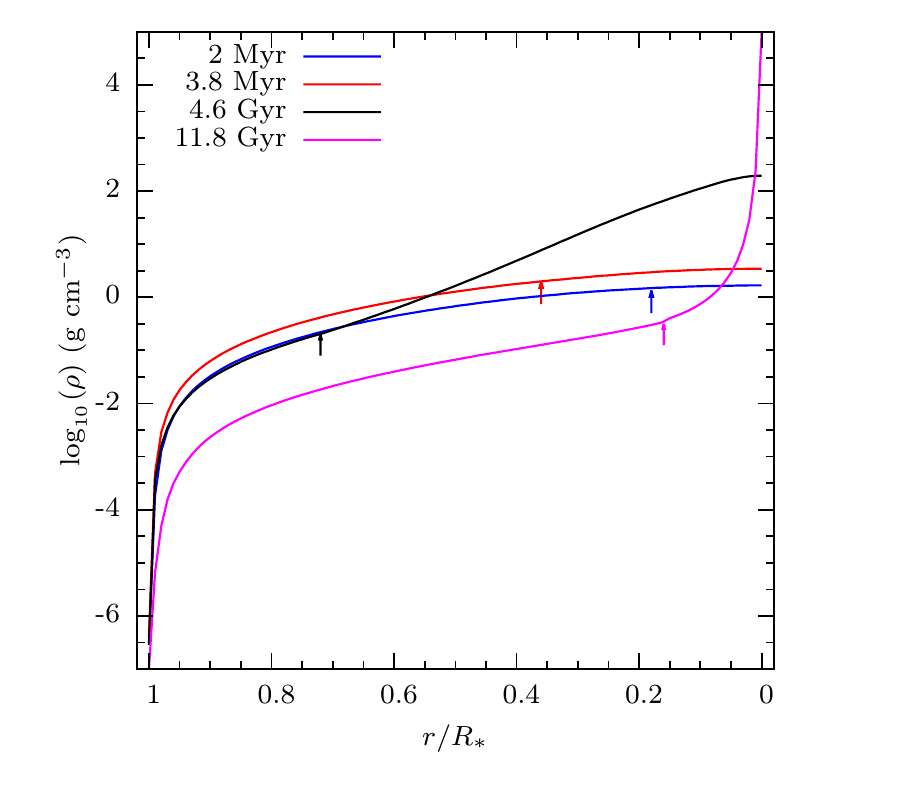}
\caption{The mass density in host star as a function of radius for different stellar ages. The arrows show the base of stellar convection zone (CZ). Blue: host star of 2 Myr, radius 1.85 $R_\odot$, the base of the CZ is at 0.18 $R_*$. Red: host star of 3.8 Myr, 1.5 $R_\odot$, the base of the CZ is at 0.36 $R_*$. Black: host star of 4.6 Gyr, 1.0 $R_\odot$, the base of the CZ is at 0.72 $R_*$. Purple: host star of 11.8 Gyr, 3.5 $R_\odot$, the base of the CZ is at 0.16 $R_*$.}
\label{rho-star}
\end{center}
\end{figure}

\begin{figure*}
\includegraphics[width=0.98\textwidth]{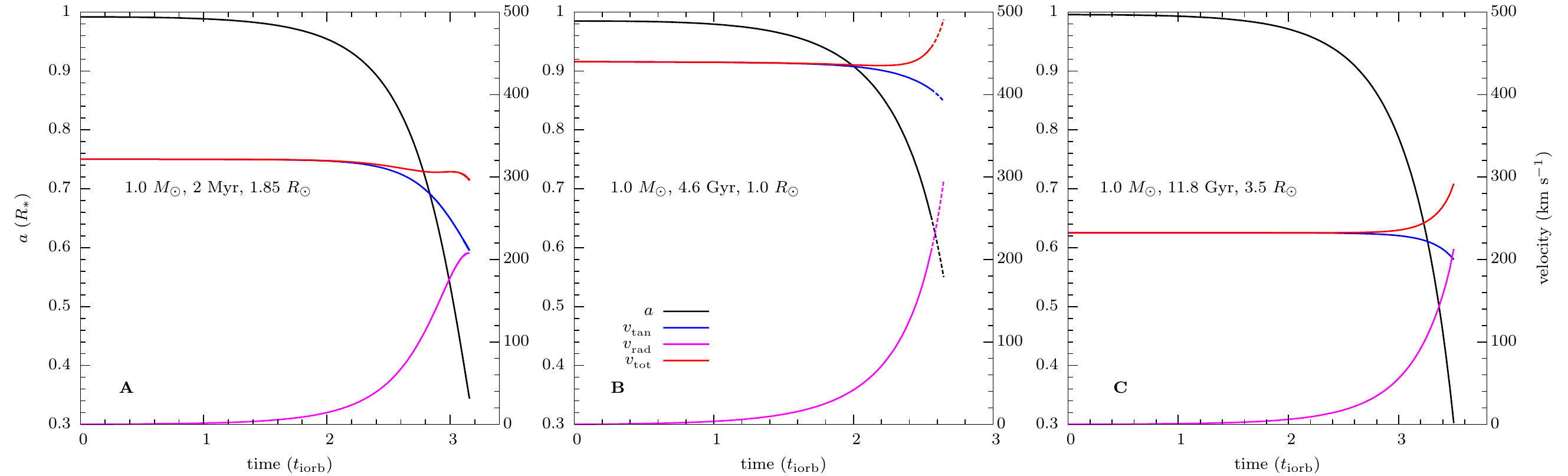}
\caption{Orbit under drag by ram pressure of a 20 $M_\oplus$ iron planet model in a convective envelope of a 1 $M_\odot$ star (solar metallicity) with different ages. The X-axis is the evolution time (in units of initial orbital period $t_{\rm iorb}$) during the planet spiral-in. The initial orbital periods from the panel A to the panel C are about $3 \times 10^3$ s, $8 \times 10^3$ s and $2 \times 10^4$ s, respectively. The black lines are the distance $a$ from the planet to the center of the host star. The red line is the velocity of the planet during its spiral-in, the blue and violet lines its tangential and radial components, respectively. Dashed extensions of the lines show the paths of the fragments beyond disruption [cf. Section \ref{debris}]. The stellar parameters are shown in the panels (see also Figure\ \ref{rho-star}).}
\label{spiral-velocity}
\end{figure*}

\subsection{Heating of the stellar envelope}
The effect of the planet on the structure of the its host has not been considered in the calculations since we expect its effect on the spiral-in process to be limited. The amount of orbital energy a Jupiter mass planet dissipates in the envelope the star can have an effect in principle, since it heats and expands the envelope. The heating has taken place in layers the planet has traveled through; deeper layers, however, are still unaffected. To lowest order in hydrostatic balance (in plane-parallel approximation: exactly), the gas pressure depends only on the weight of the mass above it. The result is that, at the depth where the planet orbits, the pressure of the stellar envelope changes only by a small amount by the heating that has gone on above it.

\section{Results}
\label{results}

\subsection{Dependence on planetary composition}
\label{compostion}
The parameters of the host star are the same in the three cases studied, a 1 $M_\odot$  pre-main sequence star with solar metallicity, radius 1.5 $R_\odot$, and age 3.8 Myr. For the host star studied here, the initial orbital period is about $5 \times 10^3$ s. The planets are a 20 $M_\oplus$ iron planet, 20 $M_\oplus$ rocky planet and 1 $M_J$ gas giant. 

\subsubsection{A 20 $M_\oplus$ Iron planet}
\label{iron}
The left panel in Figure\ \ref{deformation-iron} shows the evolution of the planet's orbit in the young host star. For the first 1.5 $t_{\rm iorb}$
the tangential velocity $\mathrm{v_{tan}}$ is almost constant, as the deceleration by ram pressure $\mathrm{P_{ram}}$ is tiny due to the low density of the upper stellar envelope (see Figure\ \ref{rho-star}). 
The velocity $\mathrm{v_{tot}}$ is still dominated by its tangential component $\mathrm{v_{tan}}$, the ram pressure $\mathrm{P_{ram}}$ is much lower than the central pressure $\mathrm{P_{cen}}$ of the planet (middle panel), and the distortion of the planet is small. 

As the planet plunges deeper into the stellar envelope (for example, $a=0.9 \ R_*$ in left panel of Figure\ \ref{deformation-iron}), the ram pressure $\mathrm{P_{ram}}$ is large enough to reduce the tangential velocities $\mathrm{v_{tan}}$ and to deform the planet as shown in the right panel in Figure\ \ref{deformation-iron}. The radial velocity increases dramatically, and the planet descends quickly into the envelope along a more radial path. The front side of the planet is compressed significantly as the ram pressure $\mathrm{P_{ram}}$ approaches the central pressure $\mathrm{P_{cen}}$ of the planet. 

The uncertainty in the onset of disruption is parame\-terized with the dimensionless coefficient $f$ in Equation (\ref{fd2}), in the following called \emph {disruption factor}. Assuming  $f=1$, the condition for disruption is met when the orbit has shrunk to about 0.6 $R_*$ (middle panel of Figure~\ref{deformation-iron}). The velocity $\mathrm{v_{tot}}$ is comparable to the initial orbital velocity (left panel). Its  radial component increases to about 200 $\mathrm{km \ s^{-1}}$. The ratio of the central pressure $\mathrm{P_{cen}}$ and the sum of $P_{\mathrm{ram}}$ and $P_{\mathrm{ext}}$ is about 1.1 at this point (middle panel). For comparison, the base of the CZ of the host star is at 0.36 $R_*$ (see also Figure\ \ref{rho-star}). This model predicts that a 20 iron $M_\oplus$ planet is disrupted within the CZ of this PMS star. The dashed extensions in Figure~\ref{deformation-iron} shows initial path the fragments would follow after disruption (see discussion in Section \ref{debris}). 

\subsubsection{A 20 $M_\oplus$ Rocky planet}
\label{rock}
The path of a 20 $M_\oplus$ rocky planet (Figure\ \ref{deformation-rock}) is similar to that of the iron planet, but it disrupts earlier. For the same mass the cross section of rocky planet is larger, and the deceleration by ram pressure correspondingly faster. The rocky planet is also more compressed and more flattened than the iron planet (right panel in Figure\ \ref{deformation-rock}).

Disruption of the rocky planet, for an assumed factor $f=1$,  occurs at $a \approx 0.86 \ R_*$. At this time, the velocity is still dominated by its tangential component during the spiral-in process, the radial component is about 70 $\mathrm{km \ s^{-1}}$.

\subsubsection{A 1 $M_J$ Giant Planet}
\label{gas}
The spiral in of the gas planet (Figure\ \ref{deformation-giant}) takes only about 0.65 $t_{\rm iorb}$. The 20 rocky $M_\oplus$ spirals in more slowly than the gas giant, but the gas giant can penetrate deeper into the envelope before disrupting. The high compressibility of the gas planet increases its density contrast with the stellar envelope, increasing its survivability as we discussed in Section \ref{compress}. The right panel of Figure\ \ref{deformation-giant} also shows the effect of the gas planet's higher compressibility.  

Disruption of the gas planet, for $f=1$, occurs at $a \approx 0.8 \ R_*$. At this time, the velocity is dominated by the tangential velocity during the spiral-in process, similar to the 20 $M_\oplus$ rocky planet,  the radial velocity is about 100 $\mathrm{km \ s^{-1}}$. 

\subsection{1 $\mathrm{M_\odot}$ host at different ages}
\label{age}
  
Here we investigate the influence of the age of the host star. The orbital velocity as the planet enters the host star is determined by stellar radius. The ram pressure, $\mathrm{P_{ram} = \rho_{ext} v^2}$, is significantly affected by the stellar structure. Figure\ \ref{rho-star} shows the density profiles for a 1 $M_\odot$ star of solar metallicity at different ages. Young PMS stars (blue and red lines) have very thick convective envelopes compared with a star of solar age (black line). 

An example of the effect of the stellar structure on the orbital evolution is shown in Figure\ \ref{spiral-velocity}. It shows the orbit of a 20 $M_\oplus$ iron planet for three different ages. Compared with the present Sun (4.6 Gyr, the base of the CZ is at 0.72 $R_*$, panel B), the orbital decay time is prolonged on the PMS (3.8 Myr, the base of the CZ is at 0.18 $R_*$, panel A) and on the post main sequence (11.8 Gyr, the base of the CZ is at 0.16 $R_*$, panel C). The orbital decay for the case of 11.8 Gyr takes 10 times as much as the case of Sun's age. In units of the planet's initial orbital period, the orbital decay time is almost same for the three cases. The planet in a solar age host (panel B) crosses the base of the CZ, with disruption  at $a \approx 0.65 \ R_*$ (for $f=1$). The planets in the case of panel A (2 Myr host star) and C (11.8 Gyr host star) penetrate deeper into the host star ($a < 0.4 \ R_*$).

\subsection{Disruption depth as a function of planetary mass}
\label{mass-dis}

\begin{figure}
\begin{center}
\includegraphics[width=0.55\textwidth]{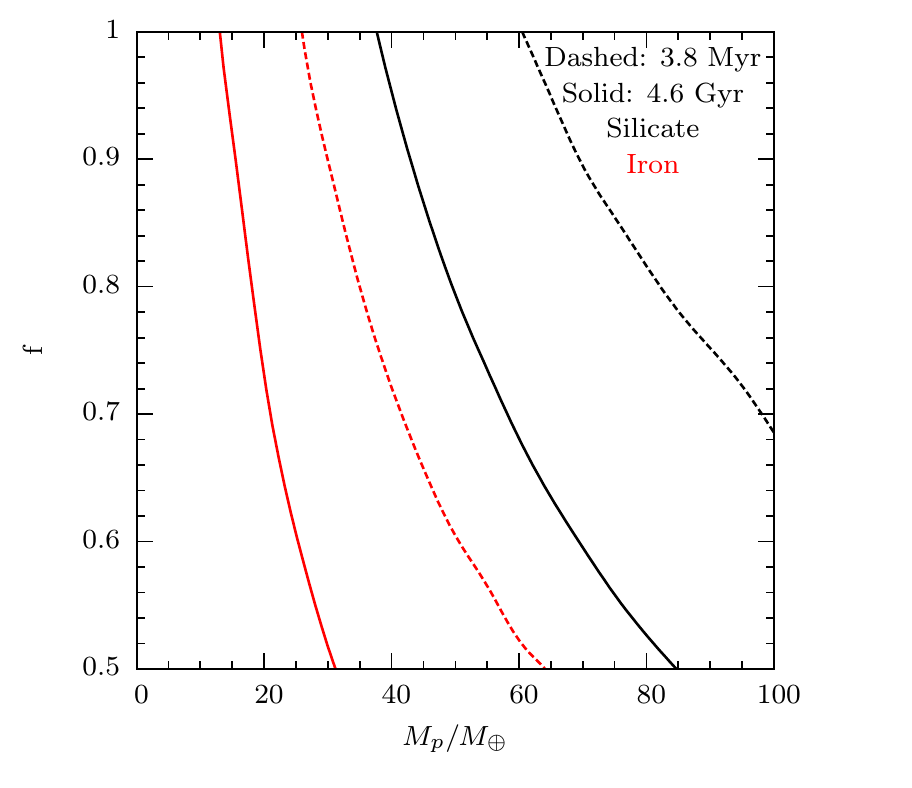} 
\caption{The minimum disruption factor $f$ required for disruption of the planet at the base of the CZ, as a function of the mass of planet. Black lines: rocky planets. Red lines: iron planets. Host star of $1\,M_\odot$, at age 3.8 Myr (dashed) and 4.6 Gyr (solid).} 
\label{bcz-dis}
\end{center}
\end{figure}

The metallicity of the stellar radiation zone will be enhanced directly if an Earth-like planet dissolves below the base of the CZ. The condition for disruption of planets in our model is a rough estimation [Equation (\ref{fd2})]. Figure\ \ref{bcz-dis} shows the minimum disruption factor $f$ required for  the planet to disrupt at the base of the stellar CZ, as a function of planetary mass. Figure\ \ref{bcz-dis} also shows that a planet (rocky or iron) needs a higher mass to survive to the base of the CZ in a PMS star than in a star of main sequence age. 

\begin{figure}
\includegraphics[width=0.55\textwidth]{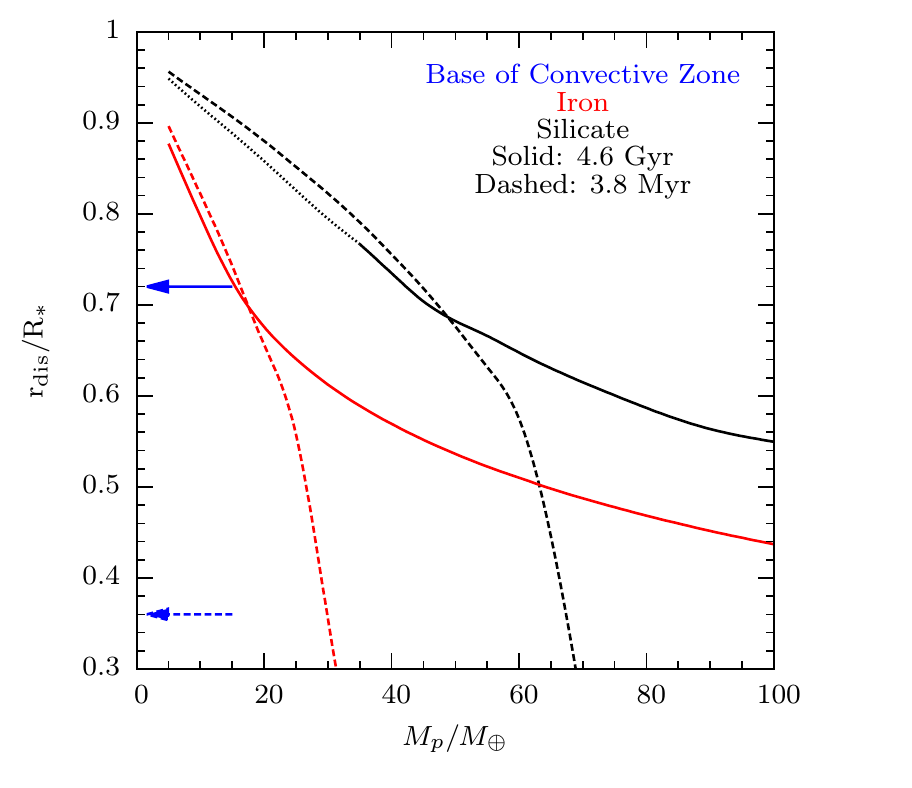}
\caption{The radial position of disruption of a planet  $\mathrm{r_{dis}}$ as a function of planet's mass for a disruption factor $f=0.9$. The blue arrows show the base of the convective envelope. Black lines: rocky planets. Red lines: iron planets. Solid lines: host star of $1\,M_\odot$, at age 4.6 Gyr. Dashed lines: host star of $1\,M_\odot$, at age 3.8 Myr. The dotted line denotes the planets that experience Roche lobe overflow before touching the surface of the $1\,M_\odot$ main sequence host star with 4.6 Gyr.} 
\label{rdis-mass}
\end{figure}

Figure \ref{rdis-mass} shows the radial position of disruption $\mathrm{r_{dis}}$ as a function of planet's mass for an assumed disruption factor $f=0.9$. The rocky planets with low mass will experience the Roche lobe overflow before touching the surface of the $1\,M_\odot$ main sequence host star with 4.6 Gyr as the dotted line shown in Figure \ref{rdis-mass} (see also Jia \& Spruit 2017). Figure \ref{rdis-mass} also shows that the planet always needs a higher mass to cross the base of the CZ before disruption for 3.8 Myr star than the Sun-like star (also from Figure\ 7). Planets of modest mass are likely to be disrupted before reaching the base of the CZ for PMS, but can survive to the base of the CZ when they enter a main sequence star.

\section{The fate of the debris}
\label{debris}

After the planet has been split by a global process (Section \ref{disruption}), the disruption process repeats itself on smaller scales until  ram pressure has reduced the velocity of the fragments below the minimum disruption size, given by Equation (\ref{tdis}).  To be determined is the development of these fragments, and where in the star they are likely to end up.

If the drag time is longer than the disruption time scale [Equation\ (\ref{dragdis})] the fragments continue along the planet's orbital path, in the opposite case the fragmentation takes place during their descent under the star's acceleration of gravity. In the following we illustrate the process for the first case,  which applies especially to  the case of a main sequence host.

\subsection{The fragmentation process}
\label{fragm}
After disruption of the planet at the depth where condition [Equation (\ref{fd2})] is first met, the mean gravitational binding energy per unit mass of the parts (scaling as $M_p/R_p$) is lower than before, having been reduced by the work done by ram pressure. As a result the condition for disruption continues to be satisfied. As long as the time to lose angular momentum by drag [Equation (\ref{dragd})] is longer than the disruption time [Equation (\ref{tdis})], the planet orbits at an approximately constant depth. The disruption time scale then does not change much since it is determined by the mean density of the fragments. To the extent that the compression of the planet and its fragments is dominated by pressure equilibrium with the ram pressure (cf. Section \ref{compress}), $\bar{\rho}_{\rm p}$ can be approximated as constant during the fragmentation process, until the point where drag begins to slow the fragments down and they start descending to deeper layers where they feel an increasing external pressure. 

We model the fragmentation as a continuous process taking place on the (initial) disruption time scale. The mass $M_{\rm f}$ of the fragments is then a function of time governed by $\rd M_{\rm f}/\rd t=-M_{\rm f}/t_{\rm d}$, so with $t_{\rm d}$ constant, 
\beq M_{\rm f}=M_0\,\exp \left(-{t \over t_{\rm d}}\right)\quad {\rm and}\quad R_{\rm f}=R_0\,\exp \left(-{t \over 3t_{\rm d}} \right),\label{mt} \eeq
where $R_{\rm f}$ is the size of the fragments, $M_0 = M_{\rm p}$ is the initial mass, $R_0 = R_{\rm p}$ is the initial radius. The deceleration of the fragments is given by the drag:
\beq {1\over 2}\rho_{\rm ext} \pv_{\rm f}^2\pi R_{\rm f}^2=-M_{\rm f}\,{\rd \pv_{\rm f} \over \rd t},\label{eqmof}\eeq
where the cross section to drag has been taken as $\pi R_{\rm f }^2$/2.  Approximating the fragments as spherical, $M_{\rm f}=\frac{4}{3}\pi \bar\rho_{\rm p} R_{\rm f}^3$, Equation \ (\ref{eqmof}) yields an equation for the velocity $\pv_{\rm f}$ which can be integrated with the result
\beq \pv_{\rm f}=\pv_0 \left\{c_0 \left[\exp\left({t \over 3t_{\rm d}} \right) -1 \right] +1 \right\}^{-1},\label{vel}\eeq
where $\pv_0 = \sqrt{GM_{\rm d}/r_{\rm d}}$ is the initial velocity and the dimensionless constant $c_0$ is
\beq c_0={9\over 8}{\rho_{\rm ext}\over{\bar\rho}_p}{t_{\rm d}\pv_0\over R_0}.\eeq
Taking $\rho_{\rm ext}/{\bar\rho}_p$ at the depth where disruption starts from Equation (\ref{fd2}) and combining Equations (\ref{tdis1}) and (\ref{torb}) this yields 
\beq c_0=\left(\frac{M_{\rm p}}{M_{\rm d}} \frac{r_{\rm d}}{R_{\rm p}} \right)^{\frac{1}{2}}=\left({{{\bar\rho}_p} \over {\bar\rho}_{\rm d}}\right)^{1 \over 2} {R_{\rm p}\over r_{\rm d}},\eeq
where the factor 9/8 has been ignored. Take as the nominal end of the orbiting phase the time $t_1$ when drag has reduced the velocity by a factor 2.  This yields with Equation (\ref{vel}):
\beq t_1=3t_{\rm d}\ln\left({1 \over c_0}+1\right)=3t_{\rm orb} \left({\bar\rho_{\rm d}\over{\bar\rho}_p}\right)^{1\over2}\ln\left({1 \over c_0}+1\right).\eeq
The mass and size of the fragments at this time follow from Equation (\ref{mt}). As an example, take a planet of Jupiter's size and mass spiraling into a PMS star of mass and radius $1 M_\odot$, $2 R_\odot$, so $c_0\approx 0.14$, then ${\bar\rho}_p/\bar\rho_{\rm d}\approx 8$, and $t_1\approx 3\, t_{\rm orb}$. The  mass and radius  of the fragments are $0.002\, M_{\rm p}$, $0.13\, R_{\rm p}$ at this time. 

\subsection{Descent}
When drag has significantly reduced the azimuthal momentum of the fragments, the orbit changes to  a descent on a more radial path. The velocity of descent $\pv_{\rm d}$ is now determined by a balance between the drag force and star's acceleration of gravity $g_{\rm r}$ rather than the planet's inertia:
\beq {1\over 2} \pi R_{\rm f}^2 \rho_{\rm ext}\pv_{\rm d}^2=g_{\rm r} M_{\rm f}, \eeq
where in view of the large density of the fragments we have ignored the buoyancy force. Disregarding also a numerical coefficient 4/3, with the star's acceleration of gravity $g_{\rm r}$ at r ($g_{\rm r} = \sqrt{GM_{\rm r}/r^2}$), this yields 
\beq \pv_{\rm d}^2\approx{\bar\rho_{\rm f}\over\bar\rho_{\rm ext}}{R_{\rm f}\over r}{2GM_{\rm r}\over r},\eeq 
or in terms of the free-fall speed $\pv_{\rm ff}$,
\beq  {{\pv_{\rm d}} \over \pv_{\rm ff} }\approx \left({\bar\rho_{\rm f}\over\bar\rho_{\rm ext}}\right)^{1 \over 2} \left({R_{\rm f}\over r}\right)^{1 \over 2},\eeq
where $M_{\rm r}$ is the mass of the star inside the radius $r$, $\bar\rho_{\rm f}$ is the mean density of the fragments. As the fragments descend, the surrounding density $\rho_{\rm ext}$ increases. At the same time, the mean density of the fragments increases by compression in the increasing external pressure. Assume that the fragments have become small enough such that their internal pressure has become approximately uniform, and assume that their structure can be taken as approximately isentropic for the present purpose. As in Section \ref{compress}, pressure equilibrium with their environment then shows that the ratio ${\bar\rho_{\rm f}/\rho_{\rm ext}}$ is approximately independent of depth. Taking a conservative value of 10 for this ratio, and the fragment size from the example of Section \ref{fragm} then yields a velocity some 20\% of the free fall speed. Within the uncertainties involved, we conclude that the time to reach the base of the CZ including spiral-in as well as fragmentation is at most a modest multiple of the initial orbital period of the planet. On this time scale, the flows in the CZ are slow, and the envelope can be treated as essentially static. 

\subsection{Settling}

The debris is cooler than its environment; its temperature will increase by radiative exchange. The time scale for this to happen, however, becomes competitive with the spiral-in time scale only for the smallest of fragments (some 10 km for conditions at the base of the solar CZ). More importantly, even when temperature equilibrium is reached, the debris of a planet of icy, rocky or iron composition will still be about twice as dense as its environment, because its (partially ionized) mean weight per particle is at least twice that of the surrounding hydrogen dominated envelope. The buoyancy of convective upflows at this depth is tiny ($\delta\rho/\rho\sim 10^{-6}$ at the base of the solar CZ). Before convective upflows can become competitive with settling, the debris therefore has to be mixed to a million times its own volume. It will settle below the CZ well before mixing to such a degree becomes relevant.

When it arrives at the stably stratified interior, the debris can not settle to an equilibrium yet. Because of its higher mean weight per particle, it can not be simultaneously in temperature, pressure and density equilibrium. Instead, it will continue to sink slowly by a `saltfingering' process such as described in Vauclair (2004).

\section{Discussion and conclusions}
 
The goal of the calculations was to quantify the processes contributing to, or delaying, the destruction of planets spiraling into their host star. The high density of a planet, compared to conditions in the greater part of a stellar envelope allow it to survive to some depth into the envelope. The process of ablation (slow peeling of the surface) turns out to be ineffective because of the large density ratio between the planet's surface and the stellar envelope (Section \ref{ablat}). This is the case even for a gas planet, because external gas and ram pressure compress its (low entropy) atmosphere to a high density. We find that the actual disruption of the planet is likely to  take place in the form of a global deformation (`splitting'), instead of by ablation. This happens when the ram pressure of the flow facing the planet is high enough to overcome the gravitational binding energy of the planet (Section \ref{disruption}). Before disruption, ram pressure deforms the planet into a flattened shape facing the flow; a model for this shape is developed in Section \ref{distortion}.  

The calculations were done for iron, rocky and gas planets entering a host star of 1 solar mass at different ages. Only planets dense enough to have avoided disruption in a previous Roche lobe overflow phase are considered. For a main sequence host, this limits the possibilities to iron-dominated or rather massive rocky or giant planets (cf. Jia \& Spruit 2017). The radii of pre-main sequence and post-main sequence hosts can be large enough (have low enough mean density) for such direct merger. 

For some combinations of  mass and composition a planet can survive its path through the entire convective envelope, disrupting finally in the radiative interior. In this way, planets can increase the metallicity preferentially in the interior rather than the convective envelope (as usually assumed). It may not be necessary that the planet survives till the base of the convection zone, however, for such `interior pollution' to work. If a rocky, icy or iron planet instead disrupts already inside a star's convective envelope, the mean mass per particle of its debris is much higher than the surroundings. The debris is likely to settle in a layer near the base of the convection zone, instead of mixing through the convective envelope (Section \ref{debris}). It can then descend into the stable interior by a `saltfingering' process as discussed in Vauclair (2004), again yielding a higher metallicity below the base of the convection zone. A sufficient mass of rocky and or iron planet(s) polluting the interior of the Sun could explain the current discrepancy between helioseismic evidence and models of the solar interior (e.g., Asplund et al. 2009; Serenelli et al. 2009; Bergemann \& Serenelli 2014; Christensen-Dalsgaard et al. 2018).

\section*{Acknowledgements} 

We would like to thank Achim Weiss for the stellar models used in this work. We also thank an anonymous referee for helpful suggestions and comments. SJ acknowledges support from the MPG-CAS Joint Doctoral Promotion Program (DPP) and Max Planck Institute for Astrophysics (MPA). SJ is also partly supported by the Natural Science Foundation of China (Grant Nos 11521303, 11390374), and the Chinese Academy of Sciences (Grant No. KJZD-EW-M06-01).

\label{lastpage}


\end{document}